\documentclass[fleqn,usenatbib]{mnras}

\usepackage{newtxtext,newtxmath}
\usepackage[T1]{fontenc}
\usepackage{pdflscape}
\usepackage{tabularx}
\usepackage{booktabs}
\usepackage{longtable}
\usepackage{array}

\DeclareRobustCommand{\VAN}[3]{#2}
\let\VANthebibliography\thebibliography
\def\thebibliography{\DeclareRobustCommand{\VAN}[3]{##3}\VANthebibliography}

\usepackage{graphicx}	% Including figure files
\usepackage{amsmath}	% Advanced maths commands
\usepackage[dvipsnames]{xcolor}

\usepackage{tikz}
\usetikzlibrary{trees}
\usetikzlibrary{arrows.meta, chains, positioning, quotes}
\usepackage{makecell}

\usepackage{xspace}
\usepackage{subcaption}

\newcommand{\erosita}{\textit{eROSITA}\xspace}
\newcommand{\srge}{\textit{SRG/eROSITA}\xspace}
\newcommand{\xmm}{\textit{\mbox{XMM-Newton}}\xspace}
\newcommand{\ergcms}{\mbox{erg s$^{-1}$ cm$^{-2}$}\xspace}
\newcommand{\lh}{Lockman Hole\xspace}
\newcommand{\nh}{N_{\rm H}}
\newcommand{\cmsqinv}{cm$^{-2}$\xspace}
\newcommand{\sqdeg}{deg$^{2}$\xspace}
\newcommand{\dense}{deg$^{-2}$\xspace}
\newcommand{\modeltwo}{\mbox{\texttt{phabs*zphabs*zpowerlw}}\xspace}
\newcommand{\clumpy}{\texttt{UXCLUMPY}\xspace}
\newcommand{\torsigma}{\texttt{TORsigma}\xspace}
\newcommand{\ctkcover}{\texttt{CTKcover}\xspace}
\newcommand{\srgeagn}{SRGe~J105348.6+573032\xspace}
\newcommand{\xmmagn}{4XMM~J105348.7+573033\xspace}

\title[\srge Reflection-dominated Compton-thick AGN candidates in Lockman Hole]{Reflection-dominated Compton-thick AGN Candidates in the \srge \lh Survey}

\author[M.I. Belvedersky et al.]{
M. I. Belvedersky$^{1,2}$,
S. D. Bykov$^{3, 4}$,
M. R. Gilfanov$^{1, 4}$\thanks{e-mail: 
gilfanov@mpa-garching.mpg.de},
P. S. Medvedev$^{1}$,
R. A. Sunyaev$^{1, 4}$ 
\\
$^{1}$Space Research Institute, Russian Academy of Sciences, Profsoyuznaya 84/32, Moscow 117997, Russia\\
$^{2}$National Research University Higher School of Economics (HSE University), Myasnitskaya 20, Moscow 101000, Russia\\
$^{3}$Kazan Federal University, Department of Astronomy and Satellite Geodesy, Kremlyovskaya 18,
Kazan 420008, Russia\\
$^{4}$Max Planck Institute for Astrophysics, Karl-Schwarzschild-Str 1, Garching b. Muenchen D-85741, Germany
}

\date{Accepted XXX. Received YYY; in original form ZZZ}

\pubyear{2024}

\begin{document}
\label{firstpage}
\pagerange{\pageref{firstpage}--\pageref{lastpage}}
\maketitle

\begin{abstract}
We search for reflection-dominated Compton-thick active galactic nuclei (CT~AGN) candidates in the \lh region using the data of \srge \lh survey. We selected sources with anomalously hard photon indices in the \mbox{0.3—8.0 keV} band, untypical for type I AGN. In particular, we required that the upper end of the 90\% error interval did not exceed a fiducial boundary of $\Gamma=1.3$. We found 291 sources which constitute a rare subpopulation among extragalactic X-ray sources detected by \erosita in the Lockman Hole field, $\approx 5\%$. These sources constitute the \erosita sample of CT~AGN candidates in the Lockman Hole field. We further divide the sources into three categories depending on the availability of reliable redshift and statistically significant detection of intrinsic absorption. We present two catalogues: the bright sample (37 sources) and the faint one (254).  We estimate the fraction and sky density of reflection-dominated CT AGN candidates. We show examples of individual spectra and use stacking analysis to search for possible redshift evolution of their properties with redshift. We analyse combined \erosita spectra of bright sources of different categories with a physically motivated spectral model \clumpy and find them fully consistent with the fits to the $\sim 1$~Msec \xmm data for one of our \mbox{reflection-dominated} CT~candidate, Type~2 galaxy SRGe~J105348.6+573032. The catalogues of CT~AGN candidates could be a good starting point for planning future studies and follow-ups at all wavelengths.
\end{abstract}


%%%%%%%%%%%%%%%%%%%%%%%%%%%%%%%%%%%%%%%%%%%%%%%%%%

%%%%%%%%%%%%%%%%% BODY OF PAPER %%%%%%%%%%%%%%%%%%

\section{Introduction}
\label{sect:intro}

An active galactic nucleus (AGN) refers to the innermost region of a galaxy that emits an exceptionally large amount of energy across the electromagnetic spectrum. AGNs are powered by the accretion of matter onto supermassive black holes located at the centres of galaxies. AGN studies give insights into the evolution of galaxies and the growth of supermassive black holes \citep{Page2004, Marconi2004, Kormendy2013, Fontanot2020}. X-ray surveys are an efficient tool for studying AGNs of various types and their populations in the Universe \citep{Brandt2015}.

In so-called obscured AGNs, the central regions are hidden from direct view by dust and gas. Such AGNs make up a significant fraction or, possibly even the majority of the AGN population  \citep{Ueda2014, Buchner2015, Ananna2019}. Dust is the dominant source of obscuration at UV–IR wavelengths, whereas gas dominates the absorption at \mbox{X-ray} energies  \citep{Hickox2018}. Heavily obscured sources may be missed even in the most sensitive X-ray surveys due to the suppression of flux in the soft X-ray range. Obscuring material is believed to be present in the form of a torus surrounding the AGN, which blocks and reprocesses radiation that originates from the accreting black hole \citep{Antonucci1993, Gilli2007, Netzer2015}.

The degree of obscuration is characterised by the hydrogen column density of gas $N_{\rm H}$, with AGNs having moderate or high column densities ($\nh~\gtrsim~10^{22}$~\cmsqinv) being referred to as obscured. Sources that have intrinsic absorption column densities $\nh~\gtrsim 10^{24}$~\cmsqinv are termed "Compton-thick" AGN (CT~AGN), as in such sources the Thomson optical depth approaches or exceeds unity \citep{Comastri2004}, making electron scatterings significant. X-ray surveys are widely used to uncover the population of obscured AGNs (including CT~AGNs), with hard band data ($>2$~keV) from Chandra \citep{Alexander2001, Alexander2003, Fiore2012, Baronchelli2017, Signorini2023}, \xmm \citep{Brusa2010, Comastri2011}, \srge \citep{Waddell2023, Waddell2024}, and harder bands ($>10$~keV) from \textit{Swift/BAT} \citep{Kawamuro2016, Ichikawa2019, Balokovic2020}, \textit{NuSTAR} \citep{Ballantyne2011, Stern2014, Brightman2015, Lansbury2017} and \textit{INTEGRAL} \citep{Sazonov2007, Malizia2012}, allowing to establish the most complete picture of AGN population \citep{Ueda2014, Buchner2015, Ananna2019, Peca2023}. For a comprehensive review of obscured AGN selection methods in other bands, see \cite{Hickox2018}.

Compton-thick AGN candidates are usually identified via reprocessed signal features, such as the Iron line and the Compton reflection continuum \citep{Guainazzi2005, Tozzi2006, Georgantopoulos2009, Done2010, Gandhi2013, Padovani2017, Buchner2021} or via observations in the high-energy ($>10$~keV) band \citep{Ricci2015, Akylas2016}. The latter is less subject to obscuration and may uncover specific spectral features such as the Compton hump \citep{Elvis2000}. Broadband spectral modelling (including soft and hard X-ray regimes) is advantageous for identifying CT~AGNs \citep{Arevalo2014, Padovani2017, Buchner2019}. Compton-thick sources constitute a significant fraction of AGNs, and their accurate counting is critical in obtaining a complete census of the population of accreting supermassive black holes \citep{Ueda2014, Buchner2015}. 

We focus on X-ray spectroscopy to search for CT~AGN candidates using data from the Lockman Hole region survey with the \erosita telescope \citep{Predehl2021} onboard the Spectrum-RG orbital X-ray observatory \citep{Sunyaev2021}. To this end, we explore reflection signatures in  X-ray spectra of \lh sources. A similar approach was used in \citet{Tozzi2006, Georgantopoulos2009}. The paper aims to systematically construct a catalogue of CT~AGN candidates that can be used as targets in future follow-ups to confirm their Compton-thick nature. 

The paper is organised as follows: first, we describe the \mbox{X-ray} data (Sect.~\ref{sect:lh_cagalogue}), then we discuss the spectral signatures of obscured AGNs relevant for this work (Sect.~\ref{sect:qual_pict}). We characterise our selection method in Sect.~\ref{sect:selection} and provide examples of spectra in Sect.~\ref{sect:spec_examples}. We explore some of our candidates by comparing them with \xmm data in Sect.~\ref{XMM_comparison}. The paper ends with a discussion of our results and a description of the obtained catalogue of CT~AGN candidates in the \lh region.

Uncertainties are quoted on a 90 per cent confidence interval, unless stated otherwise.

\section{X-ray data}

The \lh is an area located in Ursa Major that has the lowest neutral hydrogen column density on the line of sight in the entire sky \citep{Lockman1986, Dickey1990}. This condition makes the \lh a perfect region to explore extragalactic sources in general and AGNs of various types in particular. The search for obscured AGNs especially benefits from the low neutral hydrogen column density, as it minimizes the interstellar absorption of soft \mbox{X-rays}.

\subsection{\srge \lh survey}

The Lockman Hole observations took place during the performance verification phase of \srge in October 2019. The observations were conducted in a raster scanning regime, which allows for obtaining wide-field X-ray images with almost uniform sensitivity and a PSF within the image.

The footprint of the LH survey is $\approx29$~\sqdeg with the centre coordinates $\alpha=10^{\rm h}35^{\rm m}$ and $\delta=+57^{\circ}38^{\prime}$. The total survey duration is $180$~ks, mean exposure time is about $8$~ks per point. These parameters allow for achieving sensitivity $\approx3\times10^{-15}$~\ergcms in $0.5 - 2$~keV energy range. About 20\% of the survey footprint was previously observed with other X-ray telescopes such as \textit{Chandra} and \textit{XMM-Newton}.

\subsection{\srge \lh source catalogue}
\label{sect:lh_cagalogue}

The selection of candidates for highly obscured AGN is made among the sources detected by \srge \citep{Predehl2021, Sunyaev2021} during its deep survey of the Lockman Hole (LH) region. The details of the survey, \mbox{X-ray} source catalogue and procedures used for source detection and catalogue construction are presented in Gilfanov et al. (2024), in prep. The \erosita Lockman Hole source catalogue includes 6885 sources with the detection likelihood $\mathrm{DL}>10$, which approximately corresponds to the $4\sigma$ significance threshold for Gaussian distribution.

Optical identification and classification of \mbox{X-ray} sources are presented in \citet{Bykov2022, Belvedersky2022}. About $\sim1/3$ of \mbox{X-ray} sources have spectroscopic redshifts. For the remaining sources, we used photometric redshifts determined by the SRGz system \citep{Meshcheryakov2023}. The redshifts of \mbox{X-ray} sources -- both spectroscopic and photometric (SRGz), are listed in the source catalogue in Gilfanov et al. (2024) (in prep.). We apply a quality cut to the photometric redshifts as described in the following section, slightly reducing the number of available redshifts. 357 sources were excluded as they are likely Galactic based on their multiwavelength properties (see \citealt{Belvedersky2022}) and thus are not of interest in the context of this work. This way, the initial sample of sources used in this work consists of 6528 sources (all extragalactic sources in the Lockman Hole).

Source spectra were extracted for all 7 \erosita cameras using a circular aperture of 40\arcsec\ (corresponding to $\approx80$ per cent encircled energy) centred on the best-fit \mbox{X-ray} source position. To estimate the background spectrum, we used an annulus around the source with inner and outer radii of 120\arcsec\ and \mbox{250\arcsec--600\arcsec}, respectively. The outer radius was chosen so that there were at least 400 counts in the background region. The sources detected in the background region were masked using a circular aperture with a radius of \mbox{30\arcsec\--60\arcsec}, depending on their fluxes. The spectra were extracted using \texttt{srctool} task from eSASS (\erosita Science Analysis Software System, \citealt{Brunner2022}). Spectral analysis was performed with \textsc{xspec}, version 12.12 \citep{arnaud1996}. The quality of spectral fits was accessed using \mbox{W-statistic}\footnote{\url{https://heasarc.gsfc.nasa.gov/xanadu/xspec/manual}} appropriate for spectra with Poisson statistics. To avoid the well-known bias in a profile likelihood, we binned the source spectrum so that every bin in the corresponding background spectrum contains at least 5 counts. The binning was done using the \texttt{ftgrouppha} tool of the \textsc{HEASoft} package (v. 6.29).

\section{Search for Compton-thick AGN candidates}

\subsection{Qualitative picture}
\label{sect:qual_pict}

\begin{figure}
\includegraphics{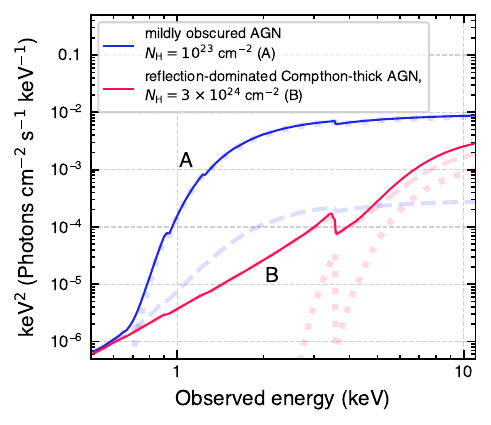}
\caption{Theoretical spectra of obscured AGN computed using \texttt{mytorus} model (zeroth-order and scattered components, no emission lines are shown) in \textsc{xspec}: (A) a mildly obscured AGN with $\nh = 10^{23}$ cm$^{-2}$ (blue lines, case (\textit{i}) in Sect.~\ref{sect:qual_pict}) and (B) Compton-thick reflection-dominated AGN with $\nh = 3\times10^{24}$ cm$^{-2}$ (pink lines, case (\textit{ii}) in Sect.~\ref{sect:qual_pict}). For both cases, solid lines show the total observed spectrum, dotted lines -- direct emission from AGN obscured with the corresponding column density, and dashed lines -- the scattered continuum (photons reflected once or more from the torus material). The inclination angle between the observer’s line of sight and the symmetry axis of the torus was assumed to be 75 degrees for both spectra. 
The redshift is $z=1$.}
\label{fig:mytorus_spectra_model}
\end{figure}

In the unification model of AGN \citep{Antonucci1993}, at shallow viewing angles, the observer registers direct emission from the accreting black hole which has passed through the torus, and reprocessed emission scattered by the torus material. In the \mbox{X-ray} band, signatures of obscured AGN can be (\textit{i}) an absorption turnover at the lower energy end of the spectrum, and/or (\textit{ii}) reflection signatures, such as a hard spectrum in the standard X-ray band, the Iron fluorescent line at $6.4-6.7$~keV with the corresponding K-edge, and the Compton hump at $\sim 20-30$~keV. Relative contributions of absorbed direct emission (\textit{i}) and reprocessed component (\textit{ii}) are determined by the viewing angle and the column density of the torus material. This behaviour is illustrated in  Fig.~\ref{fig:mytorus_spectra_model} where we show two theoretical spectra calculated using the \texttt{mytorus} model \citep{Yaqoob2012}.

In the rest frame of the source, the position of the absorption turnover in the transmitted spectrum depends on the equivalent hydrogen column density of the absorbing material. For example, for $\nh~\sim~10^{23}$~\cmsqinv,  the turnover is located at about $\sim 4$~keV.
Therefore, instruments sensitive in the hard band can detect more heavily obscured AGN compared to the telescopes operating in the soft X-ray band. However, this behaviour is modified by the redshift of the source. In the observer frame, the turnover energy is shifted to lower energies by a factor of $1+z$, where $z$ is the redshift. For this reason, identification of obscured sources (case \textit{i} above) via their low energy turnover generally requires knowledge of the source redshifts. We also note that, given the shape of the \erosita energy response (falling quickly after 2.3 keV), highly obscured sources can be plausibly detected via absorption turnover only if located at the moderate redshift of $\sim 1-2$ and/or if they are sufficiently bright.

If an AGN is heavily obscured, the low energy \mbox{X-ray} emission of the central engine might be non-detectable. In this instance, we see only emission reflected from the torus material (case \textit{ii} above) and such sources are called reflection-dominated AGNs \citep{Georgantopoulos2009, Masini2019}. The spectrum features a shallow slope in the standard X-ray band, a fluorescent Iron line, a Compton reflection hump, and other peculiarities \citep{Pounds1990, George1991, Magdziarz1995, Ross2007, Kallova2024}. Such a shallow slope cannot be produced by unsaturated comptonization in the corona \citep{Sunyaev1980} responsible for the observed X-ray spectral shape of unabsorbed AGNs. The mechanism of the \mbox{X-ray} reflection from the optically thick surface was described in \citealt{Basko1974} for a binary star system but has much in common with the spectral formation in CT AGN.

Measurement of intrinsic absorption for a reflection-dominated AGN is complicated. It requires multi-component self-consistent spectral models (e.g. \texttt{mytorus}, \citealt{Yaqoob2012}, \texttt{borus}, \citealt{Balokovic2018}, \texttt{uxclumpy},  \citealt{Buchner2019})  and a broadband spectrum of sufficient quality to accurately separate reflected and transmitted spectral components and to confirm the CT nature of the source. However, simple yet robust spectral models can be used to identify CT~AGN candidates using X-ray data, for example, by measuring the anomalously shallow photon index in the standard X-ray band \citep{Tozzi2006, Georgantopoulos2009, Georgantopoulos2013, Gandhi2013}.

In this paper, we focus on the search for reflection-dominated Compton-thick AGN. As it follows from the discussion above, the main signature of such objects in the soft X-ray band should be peculiar shallow spectra, uncharacteristic of classical type I AGN. With some caveats, candidates for such objects can also be identified without prior knowledge of source redshifts. A comprehensive search for moderately absorbed AGN among Lockman Hole sources will be performed in a follow-up publication.

\begin{figure}
 \includegraphics{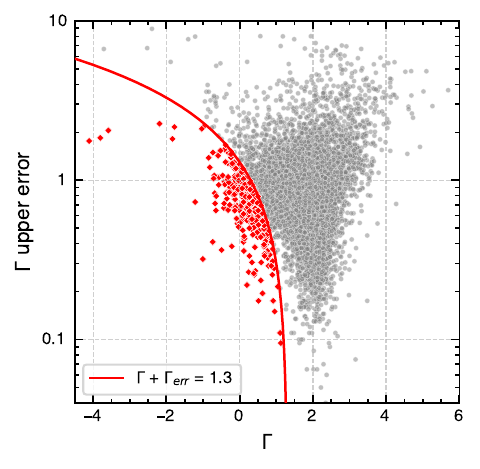}
 \caption{Best-fitting value of the photon index vs. photon index upper error (90 per cent confidence). The red curve is defined as $\Gamma + \Gamma_{\rm err} = 1.3$. Sources located below this curve and shown by red circles have a 90 per cent upper bound on their photon index lower than 1.3. A small group of 129 sources whose photon index upper bound is unconstrained were excluded from this plot.}
 \label{fig:phoind_phoind_perr}
\end{figure}

\subsection{Search for  reflection-dominated Compton-thick candidates}
\label{sect:selection}

To identify CT~AGN candidates, we employ a two-step procedure. First, we fit \mbox{X-ray} spectra of all sources from the initial sample with a power law model (\mbox{\texttt{phabs*powerlaw}} in \textsc{xspec}). The interstellar absorption is fixed at the median value for the Lockman Hole region $\nh=7\times10^{19}$~\cmsqinv \citep{HI4PICollaboration2016}. We use a conservative fiducial value of $\Gamma_0=1.3$ to separate Compton-thick AGN candidates from normal type I AGN. We use 90 per cent errors to select sources whose photon indices are lower than $\Gamma_0$: $\Gamma + \Gamma_{\rm err} < \Gamma_0$ (see Fig.~\ref{fig:phoind_phoind_perr}).  The choice of $\Gamma_0$ is motivated by the distribution of \mbox{X-ray} spectral indices with the mean $\Gamma \approx 2$ and scatter $\sigma \approx 0.2$ \citep{Nandra1994, Liu2022}. No information about the redshifts is needed at this step and, therefore, this procedure is applied to all extragalactic sources from the \erosita LH catalogue. For a small number (129) of faint sources, the 90 per cent upper bound on the photon index was unconstrained. They obviously cannot belong to the sought group of hard sources and were excluded from further analysis and from the plot in Fig.~\ref{fig:phoind_phoind_perr}. This figure shows the distribution of best-fitting photon index values and their errors. Out of 6528 sources, 291 were found to have anomalously hard spectra with photon indices lower than $\Gamma_0$ at 90 per cent confidence (red symbols). The so-constructed group of 291 sources constitutes the full sample of CT~AGN candidates.

However, such a simple approach based on the power law fit may, in some cases, misclassify mildly obscured sources as CT~AGN candidates. This is because the \mbox{low-energy} absorption turnover can mimic the shallow slope of the reflected spectrum in the standard X-ray band.

\begin{figure}
 \includegraphics{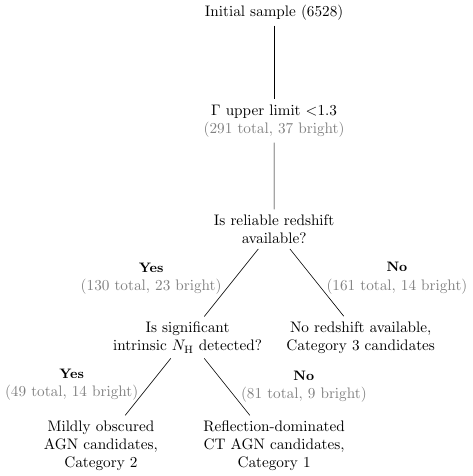}
 \caption{Classification scheme. $\Gamma$ upper limit in the second step refers to the photon index upper limit estimated using \mbox{\texttt{phabs*powerlaw}} model. Reliable redshift refers to either spectroscopic or confident photometric redshift estimation (see text for details). Bright source sample is described in Sect.~\ref{sect:bright_subsemple}.}
 \label{fig:class_scheme}
\end{figure}

To identify such cases, we fit the spectra from the initial sample of 291 CT~AGN candidates with the model which includes intrinsic absorption. We use the following \textsc{xspec} model: \modeltwo, where \texttt{phabs} describes the Galactic interstellar absorption (as in the previous step), \texttt{zphabs} describes the intrinsic absorption of the source with account for its redshift, and \texttt{zpowerlw} is a power law emission spectrum. This model requires knowledge of the source redshift, for which we used values from the \erosita LH catalogue (Gilfanov et al., 2024. in prep.). For 50 sources out of 291 spectroscopic redshifts were available. For the photometric redshifts delivered by SRGz \citep{Meshcheryakov2023}, we required that the relative redshift uncertainty taken at the 68 per cent confidence level did not exceed 30 per cent, namely \mbox{$\Delta z_{\rm phot}/(1+z_{\rm phot})<0.3$}, and that the \texttt{zConf} redshift quality parameter was larger than 0.3 (see \citealt{Meshcheryakov2023} for details). Thus, 130 sources out of 291 have sufficiently accurate redshifts (either spectroscopic or photometric) to apply the model with intrinsic absorption.

For 81 sources, no statistically significant intrinsic absorption was detected, suggesting that the data do not require absorption turnover. In combination with a $\Gamma<1.3$ spectrum, this may signal a highly obscured nature of the source and a reflection-dominated spectrum. Such sources are labelled as "Category~1" in our work. However, a word of caution is necessary -- the upper limits on intrinsic $\nh$ for many sources from Category~1 were fairly unconstraining, often in the $\sim 10^{22}-10^{23}$~cm$^{-2}$ range. Furthermore, for faint sources, the inclusion of intrinsic absorption added degeneracy to the model parameters, significantly increasing the error on the photon index. To address this issue, we introduce a bright source subsample in Sect.~\ref{sect:bright_subsemple}.

For the remaining 49 sources, labelled "Category~2" sources, the inclusion of the intrinsic absorption in the model led to a statistically significant improvement of the fit quality at a confidence level of 90 per cent -- \mbox{$\Delta$ C-stat $>2.71$} compared to the Galactic-only absorption fit. We also applied Akaike Information Criterion \citep{Akaike1974} widely used in the literature \citep[e.g.][]{Buchner2014}. As well known, AIC accounts for the difference in the number of parameters between the compared models. Expressed in terms of AIC  the Category~2 sources all have $\Delta$~AIC~$>4.71$, with the median values of $\Delta$~AIC~$>5.52$. For these 49 sources, the best-fit intrinsic column density is inconsistent with zero at the $90\%$ confidence level.

19 sources have a 90 per cent lower limit on intrinsic $\nh$ larger than $10^{22}$~\cmsqinv. No sources with intrinsic absorption above $10^{24}$~\cmsqinv were found. The best-fitting values of the photon index in the model with intrinsic absorption for these sources were found to be consistent with the canonical value of $\sim 1.9$, albeit with large uncertainties. This is quite natural as in Category~2 sources we are likely observing emission generated in the vicinity of the supermassive black hole obscured by the neutral or molecular gas with moderate $\nh$ (case \textit{i} in Sect.~\ref{sect:qual_pict}). We interpret these sources as mildly obscured (i.e. $\nh\lesssim 10^{24}$~\cmsqinv) AGNs.

The remaining 161 sources do not have sufficiently reliable redshift estimates to measure intrinsic absorption in their spectra. They are marked as "Category~3" sources. Our classification procedure is recapped in Fig.~\ref{fig:class_scheme} and category description is provided in Table~\ref{tab:category_descripiton}.

\begin{table}
    \centering
    \renewcommand{\arraystretch}{3}
    \begin{tabular}{c p{0.6\columnwidth}}
         Source category& Description\\
         \hline
         Category~1& \makecell[l]{Reflection-dominated CT~AGN candidates\\(intrinsic $\nh$ is consistent  with zero)}\\
         Category~2& \makecell[l]{Mildly obscured AGN candidates
         \\(intrinsic $\nh$ is inconsistent with zero,\\but less than $10^{24}$~\cmsqinv)} \\
         Category~3& \makecell[l]{No redshift available \\ (model without intrinsic $\nh$ is applied)} \\
    \end{tabular}
    \caption{Description of different source categories.}
    \label{tab:category_descripiton}
\end{table}

\subsection{Bright source sample}
\label{sect:bright_subsemple}

By construction, Category~1 sources should be interpreted as the best candidates for reflection-dominated CT~AGN. However, as noted above, the upper limits on the absorption column density are often unconstraining. The same can be said about the value of the photon index estimation after including the intrinsic absorption. These two effects make the classification of faint sources less robust.

To make further analysis more meaningful, we selected a bright source sub-sample from the full catalogue of 291 CT~AGN candidates. We chose the cut of 100 source counts (i.e. background subtracted) calculated with the eSASS source detection algorithm  \texttt{ERMLDET}. This approximately corresponds to the \mbox{0.3--2.3}~keV flux of \mbox{$\sim1.5\times~10^{-14}$}~\ergcms across most of the \erosita's Lockman Hole survey footprint. 37 sources pass the photon count selection and make it into the bright subsample: 9 for Category~1, 14 for Category~2 and 14 for Category~3 (see Fig.~\ref{fig:class_scheme}). They are listed in Table~\ref{tab:CT_AGN_candidates_bright}. We publish two separate catalogues -- for the bright subsample and the rest (faint sample). Together, they make up the full sample described in Sect.~\ref{sect:selection}.

\begin{figure}
 \includegraphics{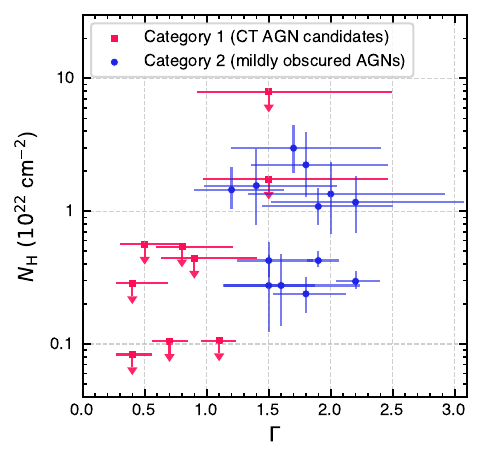}
 \caption{$\Gamma$ and intrinsic $\nh$ estimations (\modeltwo model) for the Category~1 (pink squares) and 2 (blue circles) sources from the bright source sample. Upper limits are at the 90 per cent confidence, while error bars show 1~$\sigma$ errors. One source from the Category~1 has $\Gamma = 4.6^{+2.2}_{-1.4}$, $\nh = 1.3^{+1.3}_{-0.8}$ and is not shown for the sake of visual clarity.}
 \label{fig:gamma_vs_nh_bright}
\end{figure}

In Fig.~\ref{fig:gamma_vs_nh_bright} we show $\Gamma$-$\nh$ plot for the Category 1 and 2 sources from the bright subsample. The upper limits on the intrinsic absorption are fairly constraining for almost all Category~1 sources, and their photon indices are still quite shallow. On the other hand, Category~2 sources occupy a zone of a "typical" X-ray selected mildly obscured  AGN with $\Gamma\sim2$ and $\nh\sim10^{22}$ \cmsqinv. From now on we will be referring to the sources from the bright subsample unless stated otherwise.

\begin{figure*}
 \includegraphics{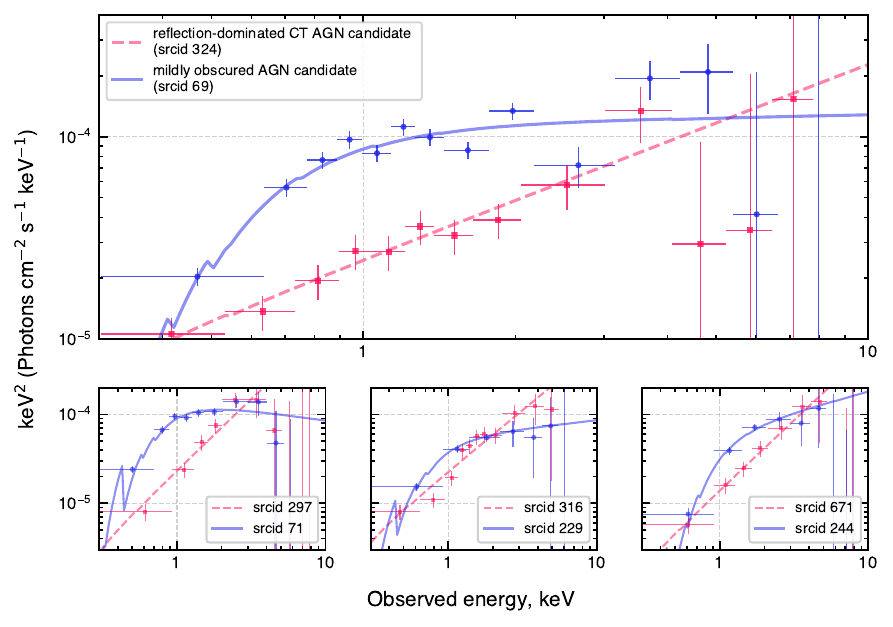}
 \caption{Blue line: a spectrum of mildly obscured ($\nh \sim 10^{22}$~\cmsqinv) source with a turnover (\modeltwo model, Category~2), red dashed line: a spectrum of CT~AGN candidate with reflection-dominated spectrum (\mbox{\texttt{phabs*powerlaw}} model, Category~1). The spectra illustrate the cases \textit{i} and \textit{ii} from the Sect.~\ref{sect:qual_pict}. Both spectra were unfolded using the model \texttt{phabs*powerlaw} with interstellar absorption fixed at $7\times10^{19}$~\cmsqinv and $\Gamma$ fixed at 2. Parameters of these sources are present in Table~\ref{tab:CT_AGN_candidates_bright}.}
\label{fig:bright_ero_spectrum_example}
\end{figure*}

\section{Spectral analysis}
\label{sect:spec_examples}

To illustrate the difference between reflection-dominated CT~AGN candidates (Category~1) and mildly obscured AGN (Category~2), we plot the spectra of the four brightest sources from each category in Fig.~\ref{fig:bright_ero_spectrum_example}. The difference in their spectral shapes is quite obvious.

\begin{figure}
    \centering
     \includegraphics[width=\columnwidth]{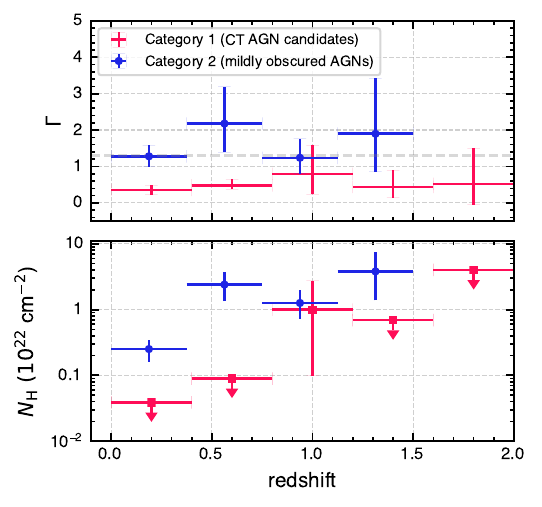}
     \caption{\textit{Upper panel:} photon index estimates obtained from \modeltwo model for simultaneously fitted spectra in different redshift bins for categories 1 and 2 (red and blue points respectively). The dashed line marks the $\Gamma=1.3$ threshold chosen in this work to separate CT~AGN candidates from other \erosita sources. \textit{Lower panel:}  $\nh$ estimations obtained from the same model, errors are for the 90\% confidence level.}
     \label{fig:stacked_z_fit_nh_phoind}
\end{figure}

For sources with spectroscopic redshifts, we searched for the possible evolution of their spectral parameters with redshift (Fig.~\ref{fig:stacked_z_fit_nh_phoind}). To this end, we grouped the sources into redshift bins. Since we performed spectral averaging in this exercise, we used all sources, not only the ones from the bright subsample. For sources of a given category and in a given redshift bin, we performed a simultaneous spectral fitting with all spectral parameters except normalisation tied between sources. The fit was performed separately for reflection-dominated CT~AGN candidates (Category~1) and mildly obscured AGNs (Category~2). The same model, \modeltwo, was used for both types of sources. We found, as expected, that photon indices of reflection-dominated CT~AGN candidates lie significantly lower than those for the sources with intrinsic absorption (Fig.~\ref{fig:stacked_z_fit_nh_phoind}, upper panel). Secondly, for CT-AGN candidates, no statistically significant intrinsic absorption was detected in all bins except the third one (the same figure, lower panel). Notably, in the first two redshift bins, where the statistical quality of combined data is sufficiently high, the upper limits on the intrinsic absorption are fairly constraining, at the level of $\sim 5\times 10^{20}$~and $\sim10^{21}$ cm$^{-2}$, and the photon indices are very shallow ($\Gamma\sim 0.5$). This is consistent with the expectation for reflection-dominated CT~AGN, although the particular value of the photon index may be affected by the presence of the soft excess in AGN spectra (see Sect.~\ref{sect:limitations} for a discussion).  In the higher redshift bins, uncertainties of the spectral parameters notably increase, due to fewer counts in analysed spectra, and the spectral fit results become less conclusive. Finally, we note that in this analysis we used a simplifying assumption that all sources of the given category in the given redshift bin have identical spectral shapes (spectral parameters except for normalisation were tied to each other). 

For the three brightest Category~1 sources with available spectroscopic redshifts (\texttt{srcid} 297, 316 and 671 from Table \ref{tab:CT_AGN_candidates_bright}), we placed upper limits on the equivalent width of the Iron line. We added a redshifted Gaussian emission line (\texttt{zgauss} in \textsc{xspec}, with a fixed energy of $6.4$ keV) to the \mbox{\texttt{phabs*powerlaw}} model and calculated 90 per cent upper limits on the equivalent width. For sources 297, 316 and 324 the upper limits are 1.9, 1.5, and 0.6 keV respectively. These numbers should be compared with the equivalent width of $\sim 1$ keV expected in the spectrum of reflected emission \citep{George1991}. In this calculation, we assumed a line width of $0.1$ keV. The upper limits reduce by $\sim 13\%$ if we assumed  intrinsically narrow lines with the width of $0.01$ keV. This exercise shows that the Iron lines (if present) are not detectable given our data. Deeper follow-up observations are needed to characterise fluorescent iron lines in these sources.

\subsection{Spectral fits with UXCLUMPY model.}
\label{sec:ero_uxclumpy}

We further explored the spectra of candidates from different categories using a physically motivated  \clumpy model. \clumpy \citep{Buchner2019} is a unification model for a clumpy obscurer in AGN. It combines a clumpy torus component with an optional "inner~ring", a \mbox{Compton-thick} reflector near the corona. The thickness of a clumpy torus is characterised by the \torsigma parameter (varying from 0 to 84 $\deg$) $-$ dispersion of the distribution of the cloud population. Parameter \ctkcover determines the covering fraction of the inner reflector and varies from 0 to 0.6.

Our attempt to fit individual \erosita spectra with \clumpy yielded unconstraining results with large uncertainties, even for the brightest sources from our sample.  We therefore applied \clumpy to combined \erosita spectra from different categories. We used sources from the bright sample having spectroscopic redshift (see Table~\ref{tab:CT_AGN_candidates_bright}), namely, four sources from Category~1 (excluding \texttt{srcid} 1430, see below) and seven sources from Category~2. The spectra were fitted simultaneously within each category (C-stat minimization in \textsc{xspec}). Galactic absorption  $\nh$$_{\rm, gal}$ was frozen at $ 7 \times 10 ^ {19}$,~\cmsqinv. Following \citet{Buchner2019}, we fixed \texttt{Ecut}  at $400$~keV, and \texttt{Theta\_inc} at $90$~$\deg$. All parameters of the model except for  line of sight $\nh$ and normalization  were linked between individual sources within each category.  The best fit parameters are presented in Table~\ref{tab:uxclumpy}. For the line of sight $\nh$ we quote mean value for each category.

\begin{figure}
    \centering
     \includegraphics[width=\columnwidth]{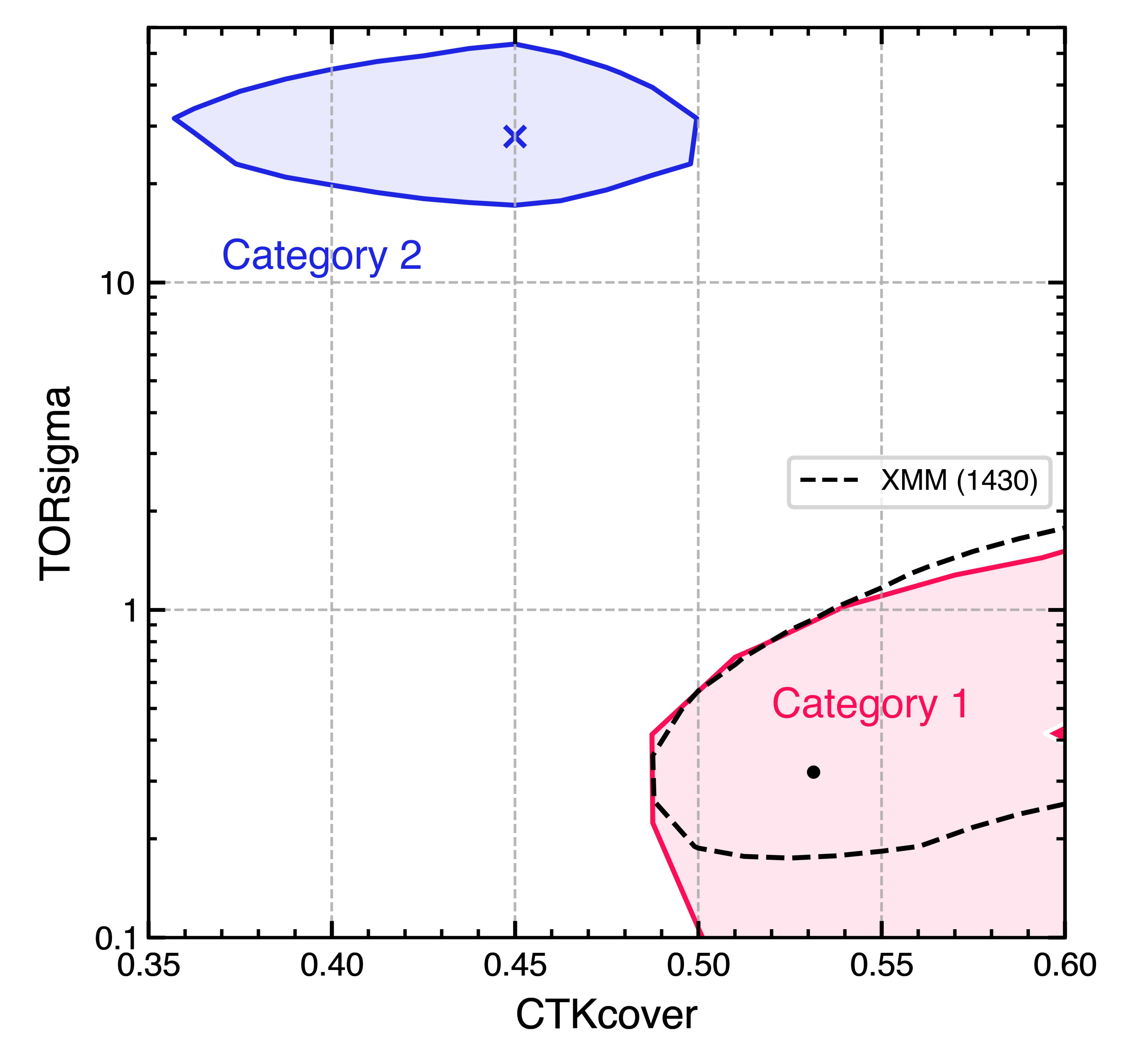}
     \caption{The covering fraction of the inner Compton thick ring \texttt{CTKover} and opening angle of the clumpy torus \texttt{TORsigma} in the \clumpy model for \erosita spectra of Category~1 (red) and Category~2 (blue) sources and XMM-Newton spectrum of \texttt{srcid} 1430 (dashed). Contours show 90\% error regions, markers show the best-fit values for each region. See Section~\ref{sec:1430_xmm} for details.}
     \label{fig:uxclumpy_erosita}
\end{figure}

As one can see from this table, candidates from different categories are clearly separated in the parameter space, as it is further illustrated in Fig.~\ref{fig:uxclumpy_erosita}. Sources from Category~1 require higher \ctkcover values  then those from Category~2. At the same time, Category~1 sources need little or no component of clumpy torus as the value of \torsigma is substantially lower for them  than for Category~2 sources.

\begin{table*}
\centering
\renewcommand{\arraystretch}{1.5}
% \begin{tabular}{p{1.7cm}>{\centering}p{1.5cm}>{\centering}p{1.4cm}c>{\centering}p{1.5cm}>{\centering\arraybackslash}p{1.1cm}cc}
\begin{tabular}{>{\centering\arraybackslash}p{1.6cm}>{\centering\arraybackslash}p{1.5cm}>{\centering\arraybackslash}p{1.5cm}>{\centering\arraybackslash}p{1.8cm}>{\centering\arraybackslash}p{1.8cm}>{\centering\arraybackslash}p{1.1cm}>{\centering\arraybackslash}p{1.8cm}}
\hline
Source type & $\nh$ [$\times 10 ^ {22}$], \cmsqinv & \texttt{PhoIndex}  & \textbf{\texttt{TORsigma}} & \textbf{\texttt{CTKcover}} & \texttt{z} (frozen)& const \\
\hline
1430 (XMM) & $1.43 \pm 0.06$ & $1.52 \pm 0.04$ & $0.31^{+1.48}_{-0.13}$ & $>0.49$ & $0.78$ & $0.09 \pm 0.08$\\
Category 1 & $2.0 \pm 0.2$ & $1.89 \pm 0.23$ & $<1.51$ & $>0.49$ & individual & $0.04 \pm 0.02$\\
Category 2 & $0.86 \pm 0.40 $ & $1.97 \pm 0.15$ & $31.65^{+21.71}_{-14.47}$ & $0.45^{+0.05}_{-0.09}$ & individual & $0.10 \pm 0.09$\\
\hline
\end{tabular}
\caption{Results of spectral fitting of \erosita and \xmm data with \clumpy model. See Sections \ref{sec:ero_uxclumpy} and \ref{sec:1430_xmm} for details. For \erosita spectra (Category 1 and 2), quoted are are mean  values of line of sight $\nh$.}
\label{tab:uxclumpy}
\end{table*}

As a   sanity check, we randomly chose five "standard" (Type I) \erosita  AGN  with $\Gamma \approx2$ (gray dots to the right of the red line in Fig.~\ref{fig:phoind_phoind_perr}) and source counts more than 300 and fitted their spectra simultaneously with \clumpy model. We found that  selected sources have \ctkcover and \torsigma consistent with zero, albeit with large uncertainties. As expected, neither Compton-thick inner ring nor absorber are needed to describe X-ray spectra of Type~1 AGN.

Thus, the \clumpy model successfully reproduced \erosita spectra of both categories of sources with compatible slope of the primary power law spectrum and with interpretable values of other key parameters of the model. In particular, we found a clear difference between Category~1 and Category~2 sources. It should be noted that the spectral fits with \clumpy model did not yield canonical values of hydrogen column density $\nh\sim 10^{24}$~\cmsqinv. This result will be discussed in section \ref{sec:discussion}.

\subsection{Comparison with the \xmm data}
\label{XMM_comparison}

For a fraction of our bright sample sources, we found counterparts of \erosita sources in the 4XMM~DR13 catalogue \citep{Webb2020}. The cross-match was performed similarly to \citealt{Bykov2022} (their Sect.~3.2). Spectroscopic redshifts were available for 2 sources from Category~1 and 3 sources from Category~2. For each XMM source, in the case of multiple \xmm observations, we selected the one with the largest exposure. We fitted XMM spectra with the \modeltwo model and used only pn detector data (for the \texttt{srcid}~1430 MOS data was also used, see below).

We used the XMM Pipeline Processing System (PPS, \citealt{Perea-Calderon2019}) data products where possible. For one object from Category~2 (\texttt{srcid}~1413) no PPS data products were available, and we used XMM observation data files (ODF) which we processed according to the standard \xmm data analysis recipies\footnote{\url{https://www.cosmos.esa.int/web/xmm-newton/sas-threads}}.

\begin{figure*}
    \includegraphics{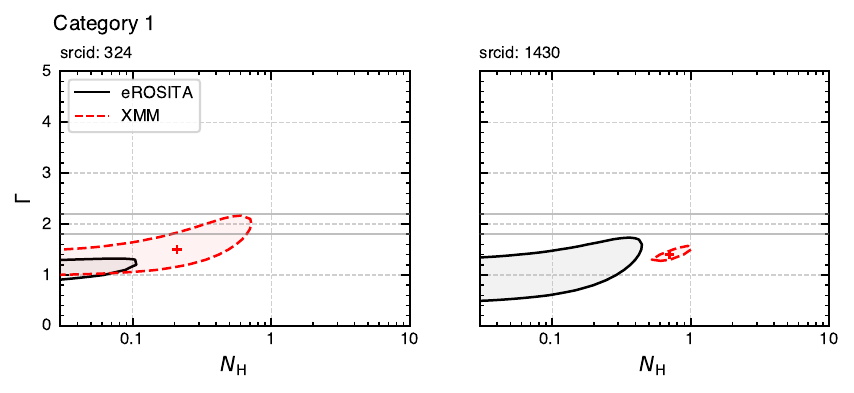}
    \includegraphics{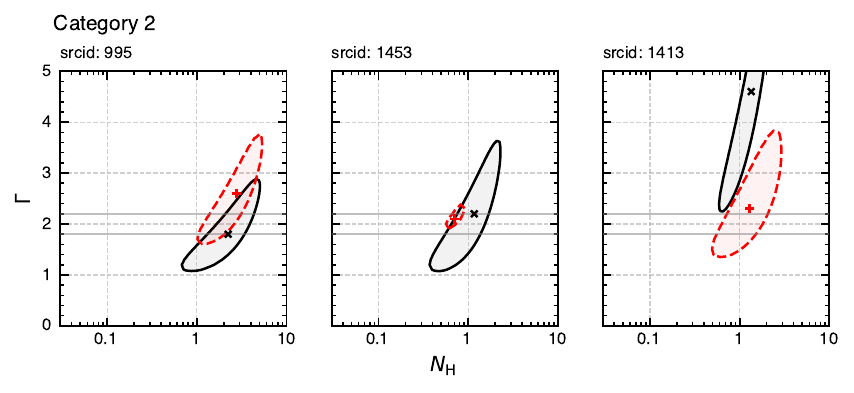}
    \caption{Comparison between the \erosita and \xmm spectral parameters. 90 per cent confidence contours of photon index $\Gamma$ vs intrinsic $\nh$ for the \erosita are plotted in solid black lines and for the XMM in dashed red lines (\modeltwo model). The first row shows examples of the Category~1 sources ($\nh$ lower limit is consistent with zero based on the \erosita data), and the second row shows a set of Category~2 sources. The stripe limited by the solid grey lines shows the photon index range $1.8 < \Gamma < 2.2$ that is typical for normal AGN \citep{Nandra1994, Liu2022}.}
    \label{fig:xmm_contours}
\end{figure*}

\begin{figure*}
    \includegraphics{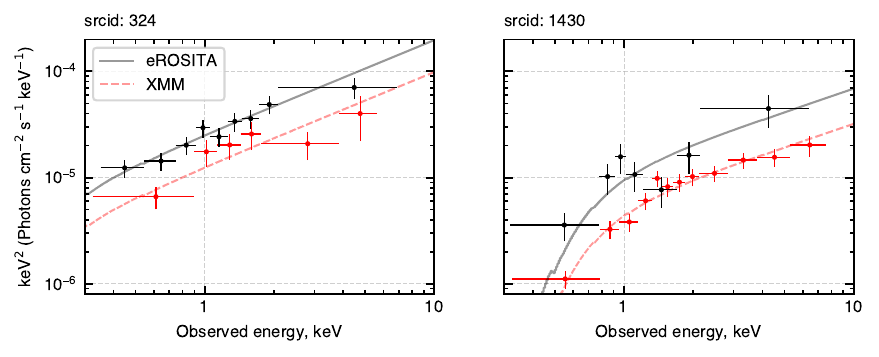}
    \caption{A comparison between (unfolded) \xmm and \erosita spectra for two brightest sources from Category~1. Both spectra were unfolded using the model \texttt{phabs*powerlaw} with interstellar absorption fixed at $7\times10^{19}$~\cmsqinv and $\Gamma$ fixed at 2.}
    \label{fig:category_1_spectra}
\end{figure*}

In Fig.~\ref{fig:xmm_contours} we compare 90 per cent confidence regions for the hydrogen column density and photon index determined from the \erosita and \xmm data. For most sources, there is good agreement between the two. \erosita data appears to be more constraining in the case of \texttt{srcid} 324. The regions do not intersect in the case of \texttt{srcid} 1430, but \xmm still shows a notably hard spectrum. The regions lie very close in case 1413. Given that we compare error regions at the 90\% some non-zero fraction of such situations is natural. We conclude that overall, there is adequate agreement between \erosita and \xmm results.
Examples of \erosita and \xmm spectra of the two brightest Category~1 sources, 324 and 1430, are shown in Fig.~\ref{fig:category_1_spectra}. 

\subsection{Detailed spectral analysis of \xmm long exposure data for SRGe~J105348.6+573032}
\label{sec:1430_xmm}

For \texttt{srcid}~1430 (\srgeagn, \xmmagn), \xmm accumulated  significant exposure time, about $\sim 10^6$ sec. For this source, we performed a detailed spectral analysis of the full set of available \xmm data using the \clumpy model. Four sets of \xmm observations of the source were performed between 2000 and 2002, with 4$-$3 observations in each set. The third set, performed between Oct.~\mbox{19$-$27} 2002, was significantly affected by the solar activity and we excluded it from our analysis. For the nine remaining observations, both the pn and MOS data were used with a total exposure of $1.04$~Msec. The combined data of all observations were fitted to the \clumpy model, the best-fit parameters are listed in Table~\ref{tab:uxclumpy}. The 90\% error region  for \torsigma  and \ctkcover is shown in Fig.~\ref{fig:uxclumpy_erosita}, along with error regions for the Category~1 and 2 combined \erosita spectra. We also fit individual sets of \xmm observations and found consistent values of the best fit parameters, albeit with large errors. From Fig.~\ref{fig:uxclumpy_erosita} and Table~\ref{tab:uxclumpy} one can see that \xmm data for \texttt{srcid}~1430 yield results which are fully consistent with those obtained from \erosita spectra of Category~1 sources. Note, that \texttt{srcid}~1430 was excluded from spectral fitting of \erosita data, shown in Fig.~\ref{fig:uxclumpy_erosita} and Table~\ref{tab:uxclumpy}. We finally note that the best fit value of the photon index in \clumpy \xmm fit is rather hard. We carefully investigated the parameter space of the model and confirmed that the minimum found is global. This behaviour might be similar, for example, to the one found for NGC~424 in \citet{Buchner2021}.

Fluorescent line of neutral iron is modelled self-consistently in \clumpy model. In order to measure its equivalent width directly from the observed spectrum, we fit the \xmm spectrum with the power law model with a superimposed Gaussian line, which energy was fixed at the redshifted position of the fluorescent line of neutral iron. Assuming intrinsically narrow line, we obtained best fit value of the equivalent width of $0.19$~keV with 90\% error interval of \mbox{$0.09-0.27$}~keV. This value corresponds to about $\sim 1/5$ of what is expected in purely reflected spectrum.

Similar to the best fit to \erosita spectra of Category~1 sources (Table~\ref{tab:uxclumpy}, Fig.~\ref{fig:uxclumpy_erosita}) we obtained rather low value of hydrogen column density in the \clumpy model, $\sim 10^{22}$~\cmsqinv. This will be discussed in the following section.

\section{Discussion}
\label{sec:discussion}

\subsection{Limitations of the analysis}
\label{sect:limitations}

The utter simplicity of the method employed in this paper to search for CT~AGN candidates allows us to apply it to faint sources with just over $\sim 20$ \erosita counts. However, this also brings about its main limitation -- a one-component spectral model, an absorbed power law, is used to characterise complex data. Real X-ray spectra of \mbox{(CT-) AGNs} are much more complicated and include several different components: transmitted emission, reflected emission with fluorescent lines and absorption edges  (see Sect.~\ref{sect:intro}), and soft excess. For this reason, the photon indices and hydrogen column density values obtained in power fits should be interpreted with a certain caution.

The picture is further complicated by the fact that the physical state of obscuring material may not be neutral, as was implicitly assumed throughout this work, but (partially) ionised.
Some AGNs show signatures of absorption of X-rays by mildly ionised optically thin gas,  called warm absorber \citep{Reynolds1995}. The material responsible for warm absorbers is linked to outflows or winds launched from broad line regions or tori \citep{Laha2014}. X-ray signatures of warm absorber systems include (blueshifted) absorption lines which require high-resolution spectroscopy to be detected \citep{Blustin2005, Kallman2019}.

The relative amplitude of reflected and direct emission may change from source to source and depends on many factors. 
The geometry of the material giving rise to the reflected emission might also not be as simple as a torus with a smooth surface and may be constrained with high-quality spectra \citep{Buchner2019}. The exact shape of the Compton hump, fluorescent lines, and the transmitted spectrum depend on the geometry  of the source and reprocessing media \citep{Nenkova2008, Buchner2021}. There is an array of spectral models with various geometries, from the solid torus to the agglomeration of blobs to simulate patchy absorbers. To name a few, \texttt{mytorus} \citep{Murphy2009, Yaqoob2012}, \texttt{borus} \citep{Balokovic2018}, \texttt{bntorus} \citep{Brightman2011}, \texttt{RXTorus} \citep{Paltani2017}, \clumpy \citep{Buchner2019} have been applied to the data from some CT~AGNs (e.g.  \citealt{Arevalo2014, Kallova2024}).

An additional component, possibly of thermal origin, is often observed in AGN at energies $\la 1$~keV, called "soft excess". It may be related to a warm corona or Comptonised emission from the accretion disc \citep{Gierlinski2004, Guainazzi2007, Done2012}. Such a component is also observed in some CT~AGNs \citep[e.g.][]{Silver2022, Peca2023, Sengupta2023}. Indeed, analysing \erosita data from the eFEDS field, \citet{Waddell2023} found that a significant portion of hard-band selected AGNs show evidence of complex absorption (with a warm absorber or an absorber with partial covering best describing the data). In addition, they found that a fraction of hard X-ray-selected AGNs possess a soft excess.  

We did not find clear statistically significant evidence of a soft excess among our CT~AGN candidates. However, the presence of a soft excess in the spectrum will effectively increase the observed photon index in the \erosita band, making sources appear softer. This may result in the loss of some of the true CT-AGNs in our selection procedure based on the photon index cut of $\Gamma<1.3$.

Complexities discussed in this subsection can potentially lead to misclassification or loss of some CT~AGNs in our selection procedure (Sect.~\ref{sect:selection}). However, a combination of the modest numbers of counts for the majority of sources and the sensitivity range of \srge, limited to the standard X-ray band, makes use of more physically motivated models such as \texttt{mytorus} unfeasible for most of the sources. As mentioned, out of 291 candidates, only 37 objects have more than 100 source counts (the typical lower limit for a meaningful X-ray spectral analysis).

All that being said, our selection procedure has picked up a small sub-population of very peculiar sources with the X-ray spectral index much harder, with sufficient statistical confidence, than canonical spectra of AGN. These sources constitute less than $\approx 5\%$ of all extragalactic sources in the Lockman Hole field detected by \srge. Their spectra are much shallower than what can be produced by unsaturated Comptonisation of soft photons \citep{Sunyaev1980} believed to be the primary mechanism of the generation of X-ray emission in AGN.
The sample of CT~AGN candidates constructed in this paper can be a good starting point for future follow-ups and broadband spectral studies.

\subsection{Comments on \clumpy spectral fit results}

In order to place our findings in the context of current understanding of CT AGN, we analysed X-ray spectra using a physically motivated spectral model. To do this, we chose the \clumpy model \citep{Buchner2019}. We fitted with this model combined \erosita spectra of Category~1 and 2 sources from the bright sample (Section~\ref{sec:ero_uxclumpy}). For comparison, we also considered a similar number of spectra of a randomly chosen standard (Type~I) AGN. We found that the \clumpy model can adequately describe both categories of \erosita sources with consistent values of the underlying power law, $\Gamma\approx 1.9$ (Table~\ref{tab:uxclumpy}) and an easily interpretable pattern of \ctkcover and \torsigma parameters, describing the covering fraction of the Compton-thick inner ring and the (angular) thickness of the clumpy absorber.  We found that while, as expected, no Compton-thick inner ring or absorber is required for the spectra of the "standard" AGN, both Category~1 and 2 sources require a Compton-thick inner ring (Fig.~\ref{fig:uxclumpy_erosita}). For Category~1 sources, larger covering fraction of the inner ring is needed but with very compact, if any, clumpy absorber outside, while for Category~2 sources the clumpy absorber seems to play a role, with significant angular dispersion of $\approx 30\degr$. This pattern was confirmed by our analysis of deep \xmm exposure of one of our Category~1 sources, Type~2 galaxy \srgeagn /\xmmagn (\texttt{srcid}~1430), for which we obtained best fit parameters fully consistent with those for \erosita spectra of Category~1 sources.

We obtained surprisingly modest values of the line of sight column density $\nh$, never reaching canonical \mbox{$\nh\sim 10^{24}$}~\cmsqinv, expected for CT~AGN. The values of line-of-sight $\nh$ are similar for \erosita spectra of Category~1 sources as well as for \texttt{srcid}~1430. The reason for such behaviour is not entirely clear. A possible explanation may be that spectral fits reported here are handicapped by the lack of \mbox{X-ray} data above $\sim 5-10$ keV (recall that the \erosita response falls sharply above $\sim2.3$ keV). Indeed, it is is well known, that broadband spectral modelling, including the hard X-ray band, is critical for analysing spectra of highly absorbed and Compton-thick AGNs \citep{Arevalo2014, Padovani2017, Buchner2019, Buchner2021}.

Alternatively, our selection algorithm might have picked up a peculiar subclass of CT AGN. \clumpy spectral fit results, taken at the face value, may suggest that in these sources the supermassive black hole is surrounded by a  inner ring of Compton-thick clouds with sufficiently large covering fraction, $\ga 0.5-0.6$. The clumpy absorber outside the inner ring is, however, quite transparent, $\nh\la 10^{23}$ \cmsqinv. In this, admittedly speculative, picture  we are observing a superposition of two components: (i) the direct emission from the vicinity of the supermassive black hole which  passed through the inner ring of clouds unaffected  and (ii) its emission reflected from the Compton-thick inner ring. Contribution of the reflected component (ii) explains shallow spectra of these sources, which can not be produced via unsaturated Comptonization, while contribution of the component (i) dilutes reflected emission and explains relative weakness of the fluorescent line of iron, observed in \texttt{srcid}~1430. In order to make further progress in uncovering the true origin of these sources, high quality broad band data is needed.

\subsection{Comments on the possible follow-up strategies}

The catalogue of potential CT~AGN candidates consists of 291 \mbox{X-ray} sources. Although we divided sources into three categories, in principle, any source in the catalogue may be a true CT~AGN.

However, we classified Category~2 sources as possibly mildly obscured AGN due to the detection of statistically significant intrinsic absorption and "normal" AGN-like slopes (albeit with large uncertainties) of their spectra. Therefore, potential X-ray follow-up should prioritise Category~1 sources, which \erosita spectra match best the expectations for CT~AGN. Since a relatively detailed spectrum is needed to establish CT nature (e.g. to resolve Iron lines, absorption edges and the Compton hump), it would make sense to observe the brightest targets first. Out of 81 sources from Category~1, 9 have more than 100 source counts (see Table~\ref{tab:CT_AGN_candidates_bright}). 

Of these 9 sources, 5 have publicly available optical spectroscopic classification (SDSS and 2MASS). Interestingly, according to available data, only one of these sources is classified as Type~2 AGN -- \texttt{srcid}~1430 which \xmm spectrum was analysed in Section~\ref{sec:1430_xmm} and also discussed above. The remaining 4 sources (\texttt{srcids} 297, 316, 671, 711) were classified as Type~1 objects, despite the fact that they have X-ray spectral characteristics similar to \texttt{srcid}~1430 and different from Type~1 AGN from Category~2. Such a discrepancy between optical classification and \mbox{X-ray} spectral properties of AGN has been known before. For example, \citealt{Guainazzi2004} described the case of Mkn~668 which features a reflection-dominated CT \mbox{X-ray}  spectrum and yet is classified as Type~1 AGN based on the presence of broad Balmer lines in its optical spectrum. They proposed several explanations for such a behaviour, including uncorrelated variability between the two bands, the possibility that young post-merger AGNs may have more compact molecular  torii  which do not effectively cover the direct view of the broad line region, and a few other possibilities.

% \textcolor{red}{\citealt{Brightman2011} and \citealt{Lacy2013} are also show that for a sufficient fraction of AGN analysed in their works the \mbox{X-ray} and optical classification contradict each other.

This may open another venue for the follow up in the optical band -- to explore spectroscopic classification of these sources further, verifying existing automated classification and making new observations when needed. We note however, that the ultimate marker for the CT nature of these objects would be follow-up X-ray observations covering hard X-ray band, above $\sim 10$ keV. Finally, we note for comparison that among 14 bright Category~2 sources, 7 have spectroscopic classification and all of them are Type~1 AGN except \texttt{srcid}~1413, which is classified as "composite" (containing both AGN activity and star formation) in \citealt{Toba2014}.

Category~3 sources do not have reliable redshift measurements. For these sources, it would make sense to select several brightest sources, e.g. with the 0.5--2 keV flux exceeding $10^{-14}$ \ergcms. For these sources, one might start with optical spectroscopy to measure their redshifts. This may be followed up with detailed X-ray spectral analysis of \erosita data and/or with deeper X-ray observations. It should also be noted that, although accurate hydrogen column density measurement is not possible without knowledge of the source redshift, a qualitative assessment of the presence (or absence) of the absorption turnover may often be possible, as illustrated in Fig.~\ref{fig:bright_ero_spectrum_example}.

Finally, we emphasise the critical importance of broadband X-ray data for any X-ray follow-up of CT~AGN candidates \citep[e.g.][]{traina2021compton}. An important role will be played by \textit{NuSTAR} \citep{Harrison2013} and \mbox{\textit{SRG/ART-XC}} \citep{Pavlinsky2021} telescopes. Their data will need to be combined with preferably quasi-contemporaneous observations in the standard X-ray band with  \xmm \citep{Jansen2001} and/or \textit{Chandra} \citep{Weisskopf2000} telescopes.

\subsection{Density of reflection dominated CT~AGN}

Based on our results, we can estimate the sky density of reflection-dominated CT~AGN candidates. Given the area of the Lockman Hole region analysed in this paper, our estimates are not very constraining. However, they may be useful in the context of further studies.

Let us assume that sources with spectra softer than the threshold used to select CT~AGN candidates (\mbox{$\Gamma - \Gamma_{\rm err} > \Gamma_{0}$}, where $\Gamma_{0} = 1.3$ is a threshold value from Section \ref{sect:selection} and $\Gamma_{\rm err}$ is the 90\% statistical uncertainty on the photon index $\Gamma$) are unlikely to be reflection-dominated CT~AGN (\mbox{non-CT} hereafter). This way, the upper limit on the fraction $f$ of reflection-dominated CT~AGN in LH will be defined as \mbox{$(N_{\rm \text{all}} - N_{\text{\rm non-CT}}) / N_{\rm \text{all}}$}, where $N_{\rm \text{all}}$ -- all extragalactic point-like sources in \lh.

The lower limit on the fraction $f$ of reflection-dominated CT~AGN in LH is defined as the number of candidates from Category~1, reflection-dominated CT~AGN candidates ($N_{\rm CT}$), divided by $N_{\rm \text{all}}$.

To estimate the sky density of reflection-dominated CT~AGN, we divided the corresponding source numbers by the sky area. The sensitivity map was considered when applying flux thresholds in counting sources. Although the sensitivity is mostly uniform throughout the entire $\approx 29$~\sqdeg Lockman Hole field, it becomes lower at the field edges. The sensitivity map was produced using the eSASS~\texttt{ersensmap} task with the detection likelihood threshold parameter \texttt{likemin} set at 10 (the same value was used to produce the \lh catalogue). 

Table \ref{tab:density_CTAGN} shows the lower and upper limit of reflection-dominated CT~AGN number $N_{\rm CT}$, sky density $\rho$ (\dense) and their fraction $f$ (in per cent) for the limiting flux $1.5 \times 10^{-14}$~\ergcms. This flux limit approximately corresponds to the typical sensitivity achieved by \erosita after 4 all-sky surveys. For the Lockman Hole survey, it approximately corresponds  to 100 source counts. With this flux limit, the area of the survey is 28.65~$\deg$, $N_{\rm all} = 1173$ and $N_{\text{\rm non-CT}} = 862$. The lower limit on the sky density  is 0.35~\dense which is consistent with predictions made for a similar limiting flux $10^{-14}$~\ergcms in \citealt{akylas2012} ($0.2 - 0.9$~\dense). We should note that the latter includes all kinds of CT~AGN, not only reflection-dominated sources.

\citealt{Ananna2019} obtained luminosity function of absorbed AGN for intrinsic column densities as high as $10^{26}$~\cmsqinv. We use their results to make an independent estimate of the expected numbers of CT AGN with $\nh>10^{25}$~\cmsqinv. To this end we use the Compton-thick ($\nh>10^{24}$~\cmsqinv) number counts from their Fig.~15 to estimate sky density of CT sources for the limiting flux of $1.5 \times 10^{-14}$~\ergcms in the  energy range of \mbox{$0.3$ -- $2.3$ keV}. In order to convert $2-7$ keV flux from their plot to our range of interest, $0.3-2.3$ keV, we used the best fit \clumpy model to the Category~1 spectra (Table~\ref{tab:uxclumpy}). We also used the assumption from \citet{Ananna2019} that the sky density of CT AGN in $\log{\nh}$ between $24-25$ and $25-26$ is equal.

We computed the sky density of CT AGNs with \mbox{$\nh > 10^{25}$}~\cmsqinv and obtained the values of $0.37 - 0.48$~\dense for the limiting flux of $1.5 \times 10^{-14}$~\ergcms in the \mbox{$0.3$ -- $2.3$ keV} energy range. This estimation a) is in good agreement with the lower limit from our measurement, Table \ref{tab:density_CTAGN} and b) it is much higher than some of the estimates obtained in previous works on the topic \citep[e.g.][]{Aird2015, Ueda2014}. We note that the advantage of X-ray luminoisty function from \citealt{Ananna2019}, which adds confidence in this number, is that it correctly predicts \textit{Chandra} COSMOS-Legacy Compton-thick number counts, see their Fig.~15 on which our estimate is also based. On the other hand, the inaccuracy of this estimate may come from the assumption of equal sky densities  of CT AGN between $\log{\nh}$ $24-25$ and $25-26$. However, this (or similar) assumption is difficult  to avoid at present as there are no or little reliable data for objects with $\nh>10^{25}$~\cmsqinv.

\begin{table}
    \centering
    \renewcommand{\arraystretch}{2.5}
    \begin{tabular}{c|cccc}
         limit & $N_{\rm CT}$ & $\rho$ & $f$ (\%) &\\
         \hline
         lower & 10 & 0.35 & $0.85 \pm 0.27$ &\\
         upper & 311 &  10.86 & $26.51 \pm 1.29$ &\\
    \end{tabular}
    \caption{Lower and upper limit estimations of CT~AGN number, sky density ($\rho$, \dense) and fraction ($f$) in per cent with the standard error. The limiting flux is $1.5 \times 10^{-14}$~\ergcms. The errors have been calculated for a binomial distribution.}
    \label{tab:density_CTAGN}
\end{table}

\section{Conclusion}

In this paper, we performed a search for reflection-dominated Compton-thick Active Galactic Nuclei candidates in the Lockman Hole region using the X-ray data from \srge telescope. Our search is based on the X-ray spectral indices of extragalactic sources detected by \erosita in this field. More specifically, we required that the upper end of the error interval for the photon index at the 90\% confidence did not exceed the fiducial value of $\Gamma_0=1.3$. Such shallow spectra are untypical for normal type I AGN and may indicate the dominance of the emission reflected from the molecular torus.

We find 291 sources with such a hard photon index. They represent a very small sub-population of extragalactic X-ray sources detected by \erosita in this region, about $\approx 5\%$. Of those 291 sources, 81 have no evidence of intrinsic absorption, 49 have intrinsic absorption and are likely moderately obscured AGNs, and 161 sources have no reliable redshift to perform the intrinsic absorption analysis. These sources form three categories which may contain bona fide CT~AGNs with reflection-dominated spectra.

We present two catalogues, the bright subsample (with source counts higher than 100, 37 sources), and the faint one (254 sources). We list their X-ray properties and estimate their fraction (among all sources) and sky density. We show examples of individual spectra and use stacking analysis to search for possible evolution of their properties with redshift. 

In order to interpret our findings in the framework of physically motivated models, we fit grouped \erosita spectra of sources in the bright sample with \clumpy model \citep{Buchner2019} and found that Category~1 and 2 sources occupy distinct regions in the parameter space of this model. Using terminology of the \clumpy model, Category~1 sources require larger solid angle subtended by the Compton-thick inner screen and much more compact clumpy absorber than Category~2 sources. This trend is also confirmed by our analysis of the ultra deep ($\sim 1$ Msec) data on one of our Category~1 sources, Type~2 galaxy \srgeagn/\xmmagn (\texttt{srcid}~1430). Its best fit and error region for \torsigma and \ctkcover parameters of the model are fully consistent with those for the Category~1 \erosita sources. However, our spectral fits yielded surprisingly moderate values the line-of-sight $\nh$ values, never reaching $\nh\sim 10^{24}$ cm$^{-2}$ typically expected for CT AGN. We tentatively proposed a  plausible interpretation of this result.

We discuss the limitations of the analysis due to the low number of counts and the implications of our inability to apply sophisticated models for spectral fitting. The catalogues of reflection-dominated CT~AGN candidates can be used for planning future studies and follow-ups, and possible strategies are discussed.

Further observations of the selected sources from our catalogue of reflection-dominated CT~AGN candidates with higher sensitivity in the hard X-ray band, first of all using \textit{NuSTAR} and \ \mbox{\textit{SRG/ART-XC}} telescopes, will permit the search for fluorescent iron line and Compton reflection continuum, establishing or refuting the highly obscured nature of the sources from our catalogue.

The Lockman Hole region presents a good opportunity for studies of obscured AGNs. It has extensive multiwavelength coverage advantageous for more effective identification of obscured sources, very low interstellar absorption for studying their soft X-ray properties, and a large enough area to discover rare objects.  The extension of the method to the \erosita all-sky survey is expected to produce many more CT~AGNs candidates.  The refinement of the method with the all-sky data and subsequent follow-up will allow for an increase in the purity and completeness of candidate selection.

\section*{Acknowledgements}

This work is based on observations with \erosita telescope onboard SRG observatory. The SRG observatory was built by Roskosmos in the interests of the Russian Academy of Sciences represented by its Space Research Institute (IKI) in the framework of the Russian Federal Space Program, with the participation of the Deutsches Zentrum für Luft- und Raumfahrt (DLR). The \srge \mbox{X-ray} telescope was built by a consortium of German Institutes led by MPE, and supported by DLR. The SRG spacecraft was designed, built, launched and is operated by the Lavochkin Association and its subcontractors. The science data are downlinked via the Deep Space Network Antennae in Bear Lakes, Ussurijsk, and Baykonur, funded by Roskosmos. The \erosita data used in this work were processed using the eSASS software system developed by the German \erosita consortium and proprietary data reduction and analysis software developed by the Russian \erosita Consortium. This work was supported by the Ministry of Science and Higher Education grant 075-15-2024-647.

This work used the data obtained with \xmm, an ESA science mission with instruments and contributions directly funded by ESA Member States and NASA.

The following software was used: \textsc{xspec}, \citep{arnaud1996}, NumPy \citep{Harris2020}, Matplotlib \citep{Hunter2007},  SciPy \citep{2020SciPy-NMeth}, pandas \citep{mckinney-proc-scipy-2010}, AstroPy \citep{astropy:2018}.

The authors would like to thank the anonymous referee for insightful and constructive comments which helped to improve the paper.

%%%%%%%%%%%%%%%%%%%%%%%%%%%%%%%%%%%%%%%%%%%%%%%%%%
\section*{Data Availability}

The catalogues presented in this article are available as online supplementary material.
\srge data on the sources published in this paper can be made available upon a reasonable request. The catalogue of reflection-dominated Compton-thick AGN candidates presented here will be made publicly available via the VizieR\footnote{\url{https://vizier.cds.unistra.fr/viz-bin/VizieR}} system after the publication of this work.

\bibliographystyle{mnras}
\bibliography{citations} % if your bibtex file is called example.bib

\newpage
\appendix
\section{The bright source catalogue}

Table \ref{tab:CT_AGN_candidates_bright} shows the bright subsample (see Sect.~\ref{sect:bright_subsemple}), sorted in decreasing X-ray flux order (within each category). Only a subset of columns is shown. The full catalogue is available in the supplementary materials and via \href{https://vizier.cds.unistra.fr/viz-bin/VizieR}{VizieR}, as well as a separate catalogue for the faint sample. The excerpt of the latter is shown in Table \ref{tab:CT_AGN_candidates_faint}.

\subsection{Catalogue discription}

Both the bright and the faint source catalogues of \srge 
reflection-dominated Compton-thick AGN candidates in the Lockman Hole will be available via the VizieR web service and in supplementary materials. The bright source catalogue contains 37 rows, the faint one -- 254. Both contain 23 columns. The columns include the source index, name and coordinates of both the \srge sources and their optical candidates taken from \cite{Bykov2022}. Each candidate has a category assigned in Sect.~\ref{sect:selection}, description of these categories is given in Table~\ref{tab:category_descripiton}. Parameters and errors are provided for two models, \mbox{\texttt{phabs*powerlaw}} and \modeltwo. Errors are quoted at the 90 per cent confidence level. \mbox{X-ray} characteristics include 0.5-2.0 keV flux with 1-sigma error and source counts. Spectroscopic redshifts for some optical counterparts were obtained in \citet{Belvedersky2022}. Photometric redshifts were derived using the SRGz system \citep{Meshcheryakov2023}.

\onecolumn
\begin{landscape}
    \fontsize{8}{12}\selectfont
    \renewcommand{\arraystretch}{1.6}
\begin{longtable}{lllcrrrrllllcl}
% \begin{tabularx}
% \caption{\srge CT AGN candidates in \lh}
\toprule
 & $^{(1)}$\texttt{srcid} & $^{(2)}$\texttt{NAME} & $^{(3)}$\texttt{category} & $^{(4)}$\texttt{$\Gamma_{\text{phabs*po}}$} & $^{(5)}$\texttt{$\Gamma$} & $^{(6)}$\texttt{$N_{\rm H}$} [$\times 10 ^ {22}$] & $^{(7)}$\texttt{RA\_X} & $^{(8)}$\texttt{DEC\_X} & $^{(9)}$\texttt{RA\_OPT} & $^{(10)}$\texttt{DEC\_OPT} & $^{(11)}$\texttt{FLUX\_X} [$\times 10 ^ {-14}$] & $^{(12)}$\texttt{redshift} \\
\midrule
1 & 297 & SRGe J103841.5+604401 & 1 & $0.40^{+0.26}_{-0.25}$ & $0.40^{+0.25}_{-0.22}$ & $< 0.44$ & 159.673 & 60.734 & 159.674 & 60.734 & $4.78 \pm 0.43$ & 0.52 (spec) \\
2 & 316 & SRGe J104108.1+562000 & 1 & $0.62^{+0.19}_{-0.20}$ & $0.70^{+0.25}_{-0.24}$ & $< 0.08$ & 160.284 & 56.333 & 160.284 & 56.333 & $4.58 \pm 0.30$ & 0.23 (spec) \\
3 & 324 & SRGe J104051.4+573439 & 1 & $1.04^{+0.21}_{-0.21}$ & $1.10^{+0.22}_{-0.24}$ & $< 0.29$ & 160.214 & 57.578 & 160.215 & 57.578 & $4.56 \pm 0.29$ & 0.73 (phot) \\
4 & 671 & SRGe J103039.8+580610 & 1 & $0.43^{+0.27}_{-0.27}$ & $0.40^{+0.48}_{-0.22}$ & $< 0.11$ & 157.666 & 58.103 & 157.665 & 58.103 & $2.87 \pm 0.26$ & 0.50 (spec) \\
5 & 711 & SRGe J104711.3+582820 & 1 & $0.39^{+0.26}_{-0.25}$ & $0.50^{+0.53}_{-0.33}$ & $< 0.57$ & 161.797 & 58.472 & 161.796 & 58.472 & $2.74 \pm 0.24$ & 0.75 (spec) \\
6 & 1246 & SRGe J105553.9+592410 & 1 & $0.71^{+0.35}_{-0.34}$ & $1.50^{+1.58}_{-0.87}$ & $< 0.54$ & 163.975 & 59.403 & 163.975 & 59.403 & $1.80 \pm 0.19$ & 0.68 (phot) \\
7 & 1267 & SRGe J104213.2+584706 & 1 & $0.79^{+0.36}_{-0.36}$ & $0.80^{+0.68}_{-0.35}$ & $< 1.75$ & 160.555 & 58.785 & 160.555 & 58.785 & $1.77 \pm 0.19$ & 0.74 (phot) \\
8 & 1430 & SRGe J105348.6+573032 & 1 & $0.86^{+0.42}_{-0.42}$ & $0.90^{+0.83}_{-0.44}$ & $< 7.93$ & 163.452 & 57.509 & 163.453 & 57.509 & $1.61 \pm 0.19$ & 0.78 (spec) \\
9 & 1660 & SRGe J103620.2+555546 & 1 & $0.71^{+0.41}_{-0.39}$ & $1.50^{+1.64}_{-0.95}$ & $< 0.11$ & 159.084 & 55.930 & 159.085 & 55.930 & $1.43 \pm 0.19$ & 1.79 (phot) \\
10 & 69 & SRGe J104130.0+551414 & 2 & $1.12^{+0.09}_{-0.09}$ & $1.90^{+0.27}_{-0.15}$ & $0.43^{+0.13}_{-0.08}$ & 160.375 & 55.237 & 160.375 & 55.237 & $13.26 \pm 0.41$ & 0.75 (phot) \\
11 & 71 & SRGe J104014.0+595732 & 2 & $1.10^{+0.11}_{-0.11}$ & $2.20^{+0.33}_{-0.26}$ & $0.30^{+0.10}_{-0.06}$ & 160.058 & 59.959 & 160.058 & 59.959 & $12.82 \pm 0.49$ & 0.24 (phot) \\
12 & 229 & SRGe J105421.3+572543 & 2 & $0.92^{+0.18}_{-0.17}$ & $1.80^{+0.53}_{-0.44}$ & $0.24^{+0.14}_{-0.11}$ & 163.589 & 57.429 & 163.588 & 57.429 & $5.73 \pm 0.33$ & 0.21 (spec) \\
13 & 244 & SRGe J105421.4+582345 & 2 & $0.52^{+0.17}_{-0.17}$ & $1.50^{+0.58}_{-0.42}$ & $0.43^{+0.26}_{-0.17}$ & 163.589 & 58.396 & 163.590 & 58.396 & $5.51 \pm 0.32$ & 0.20 (spec) \\
14 & 514 & SRGe J104336.9+581439 & 2 & $0.20^{+0.22}_{-0.22}$ & $1.20^{+0.69}_{-0.50}$ & $1.45^{+1.14}_{-0.68}$ & 160.904 & 58.244 & 160.903 & 58.244 & $3.43 \pm 0.26$ & 0.82 (spec) \\
15 & 786 & SRGe J104144.7+552957 & 2 & $0.94^{+0.28}_{-0.27}$ & $1.50^{+0.61}_{-0.60}$ & $0.28^{+0.31}_{-0.25}$ & 160.436 & 55.499 & 160.436 & 55.498 & $2.54 \pm 0.22$ & 0.73 (phot) \\
16 & 822 & SRGe J105726.4+592844 & 2 & $0.24^{+0.27}_{-0.26}$ & $1.90^{+1.00}_{-0.75}$ & $1.09^{+0.66}_{-0.48}$ & 164.360 & 59.479 & 164.361 & 59.478 & $2.47 \pm 0.23$ & 0.21 (phot) \\
17 & 995 & SRGe J104732.2+575148 & 2 & $0.79^{+0.29}_{-0.29}$ & $1.80^{+1.09}_{-0.72}$ & $2.24^{+2.81}_{-1.56}$ & 161.884 & 57.863 & 161.885 & 57.863 & $2.11 \pm 0.21$ & 1.53 (phot) \\
18 & 1010 & SRGe J104800.7+551503 & 2 & $0.68^{+0.33}_{-0.33}$ & $2.00^{+1.52}_{-1.10}$ & $1.35^{+1.66}_{-1.11}$ & 162.003 & 55.251 & 162.004 & 55.251 & $2.08 \pm 0.23$ & 0.71 (spec) \\
19 & 1036 & SRGe J104037.1+600006 & 2 & $0.77^{+0.33}_{-0.32}$ & $1.60^{+1.05}_{-0.78}$ & $0.28^{+0.33}_{-0.23}$ & 160.154 & 60.002 & 160.153 & 60.002 & $2.05 \pm 0.21$ & 0.25 (phot) \\
20 & 1413 & SRGe J104809.6+565500 & 2 & $0.41^{+0.38}_{-0.37}$ & $4.60^{+3.67}_{-2.35}$ & $1.35^{+1.32}_{-0.76}$ & 162.040 & 56.917 & 162.040 & 56.917 & $1.63 \pm 0.19$ & 0.05 (spec) \\
21 & 1453 & SRGe J104655.5+590301 & 2 & $0.67^{+0.40}_{-0.39}$ & $2.20^{+1.43}_{-1.13}$ & $1.17^{+1.12}_{-0.80}$ & 161.731 & 59.050 & 161.731 & 59.050 & $1.60 \pm 0.18$ & 0.83 (spec) \\
22 & 1466 & SRGe J104511.3+574151 & 2 & $0.27^{+0.33}_{-0.32}$ & $1.70^{+1.16}_{-0.83}$ & $2.98^{+2.44}_{-1.70}$ & 161.297 & 57.698 & 161.298 & 57.697 & $1.58 \pm 0.19$ & 1.09 (phot) \\
23 & 3252 & SRGe J104321.1+545629 & 2 & $0.68^{+0.34}_{-0.33}$ & $1.40^{+1.07}_{-0.69}$ & $1.56^{+2.28}_{-1.28}$ & 160.838 & 54.941 & 160.837 & 54.941 & $0.81 \pm 0.10$ & 1.74 (spec) \\
24 & 276 & SRGe J102200.4+561225 & 3 & $0.95^{+0.15}_{-0.15}$ & $-$ & $-$ & 155.502 & 56.207 & 155.502 & 56.207 & $4.99 \pm 0.25$ & $-$ \\
25 & 601 & SRGe J105818.3+580749 & 3 & $0.57^{+0.24}_{-0.23}$ & $-$ & $-$ & 164.576 & 58.130 & 164.576 & 58.130 & $3.10 \pm 0.26$ & $-$ \\
26 & 808 & SRGe J103143.5+573252 & 3 & $0.86^{+0.29}_{-0.29}$ & $-$ & $-$ & 157.931 & 57.548 & 157.930 & 57.548 & $2.49 \pm 0.23$ & $-$ \\
27 & 830 & SRGe J103120.5+554142 & 3 & $0.90^{+0.25}_{-0.24}$ & $-$ & $-$ & 157.835 & 55.695 & 157.835 & 55.694 & $2.46 \pm 0.21$ & $-$ \\
28 & 942 & SRGe J105101.6+564136 & 3 & $0.90^{+0.31}_{-0.30}$ & $-$ & $-$ & 162.757 & 56.693 & 162.757 & 56.694 & $2.21 \pm 0.21$ & $-$ \\
29 & 1077 & SRGe J104231.7+570006 & 3 & $0.59^{+0.31}_{-0.30}$ & $-$ & $-$ & 160.632 & 57.002 & 160.634 & 57.002 & $1.99 \pm 0.20$ & $-$ \\
30 & 1155 & SRGe J104222.7+602647 & 3 & $0.78^{+0.35}_{-0.34}$ & $-$ & $-$ & 160.595 & 60.446 & 160.595 & 60.446 & $1.90 \pm 0.20$ & $-$ \\
31 & 1225 & SRGe J104614.7+572426 & 3 & $0.34^{+0.30}_{-0.30}$ & $-$ & $-$ & 161.561 & 57.407 & 161.561 & 57.407 & $1.82 \pm 0.20$ & $-$ \\
32 & 1319 & SRGe J105200.5+585350 & 3 & $0.70^{+0.38}_{-0.36}$ & $-$ & $-$ & 163.002 & 58.897 & 163.000 & 58.897 & $1.72 \pm 0.19$ & $-$ \\
33 & 1381 & SRGe J103602.3+563541 & 3 & $0.91^{+0.37}_{-0.35}$ & $-$ & $-$ & 159.010 & 56.595 & 159.011 & 56.595 & $1.65 \pm 0.18$ & $-$ \\
34 & 1454 & SRGe J104110.7+571341 & 3 & $0.76^{+0.41}_{-0.41}$ & $-$ & $-$ & 160.295 & 57.228 & 160.296 & 57.228 & $1.60 \pm 0.19$ & $-$ \\
35 & 1520 & SRGe J102305.5+564109 & 3 & $0.76^{+0.36}_{-0.36}$ & $-$ & $-$ & 155.773 & 56.686 & 155.773 & 56.686 & $1.53 \pm 0.16$ & $-$ \\
36 & 1662 & SRGe J103755.0+554548 & 3 & $0.12^{+0.36}_{-0.35}$ & $-$ & $-$ & 159.479 & 55.763 & 159.479 & 55.763 & $1.43 \pm 0.17$ & $-$ \\
37 & 1673 & SRGe J104642.2+545425 & 3 & $0.82^{+0.34}_{-0.34}$ & $-$ & $-$ & 161.676 & 54.907 & 161.675 & 54.908 & $1.42 \pm 0.14$ & $-$ \\
\bottomrule
\caption{\srge CT AGN candidates in \lh, bright sample. (1) -- source ID from the eROSITA Lockman Hole catalogue, (2) -- eROSITA source name, (3) -- candidate's category, (4) -- photon index derived from the \mbox{\texttt{phabs*powerlaw}} model, (5, 6) -- photon index and absorbing column density (in \cmsqinv) for \texttt{phabs~*~zphabs~*~powerlw} model, (7-10) -- X-ray and optical coordinates of the source in degrees, (11) -- X-ray flux in \ergcms for 0.5-2.0 keV range, (12) -- redshift (spec-z or photo-z). A full table with additional columns is available in the supplementary materials.}
% \end{tabularx}
\end{longtable}
    \label{tab:CT_AGN_candidates_bright}
\end{landscape}

\clearpage
\begin{landscape}
    \fontsize{8}{12}\selectfont
    \renewcommand{\arraystretch}{1.6}
\begin{longtable}{lllcrrrrllllcl}
\toprule
 & $^{(1)}$\texttt{srcid} & $^{(2)}$\texttt{NAME} & $^{(3)}$\texttt{category} & $^{(4)}$\texttt{$\Gamma_{\text{phabs*po}}$} & $^{(5)}$\texttt{$\Gamma$} & $^{(6)}$\texttt{$N_{\rm H}$} [$\times 10 ^ {22}$] & $^{(7)}$\texttt{RA\_X} & $^{(8)}$\texttt{DEC\_X} & $^{(9)}$\texttt{RA\_OPT} & $^{(10)}$\texttt{DEC\_OPT} & $^{(11)}$\texttt{FLUX\_X} [$\times 10 ^ {-14}$] & $^{(12)}$\texttt{redshift} \\
\midrule
1 & 519 & SRGe J105626.8+563240 & 1 & $0.67^{+0.52}_{-0.51}$ & $1.00^{+1.29}_{-0.78}$ & $< 0.64$ & 164.112 & 56.544 & 164.113 & 56.545 & $3.40 \pm 0.53$ & 0.47 (phot) \\
2 & 1420 & SRGe J103741.3+595048 & 1 & $0.62^{+0.45}_{-0.45}$ & $0.70^{+0.36}_{-0.44}$ & $< 0.05$ & 159.422 & 59.847 & 159.423 & 59.846 & $1.62 \pm 0.21$ & 0.09 (spec) \\
3 & 1587 & SRGe J103506.8+562849 & 1 & $0.70^{+0.42}_{-0.40}$ & $0.90^{+0.91}_{-0.56}$ & $< 3.84$ & 158.778 & 56.480 & 158.779 & 56.480 & $1.48 \pm 0.18$ & 0.46 (spec) \\
4 & 2224 & SRGe J105528.6+573122 & 1 & $0.76^{+0.46}_{-0.45}$ & $1.50^{+3.02}_{-1.04}$ & $< 0.54$ & 163.869 & 57.523 & 163.868 & 57.523 & $1.13 \pm 0.16$ & 0.50 (phot) \\
5 & 2512 & SRGe J104213.6+575101 & 1 & $0.36^{+0.45}_{-0.44}$ & $0.90^{+1.27}_{-0.85}$ & $< 2.85$ & 160.557 & 57.850 & 160.556 & 57.850 & $1.02 \pm 0.16$ & 1.13 (spec) \\
$\vdots$ & $\vdots$ & $\vdots$ & $\vdots$ & $\vdots$ & $\vdots$ & $\vdots$ & $\vdots$ & $\vdots$ & $\vdots$ & $\vdots$ & $\vdots$ & $\vdots$ \\
82 & 2365 & SRGe J103544.9+571117 & 2 & $0.52^{+0.53}_{-0.51}$ & $1.50^{+1.71}_{-1.11}$ & $1.68^{+2.94}_{-1.61}$ & 158.937 & 57.188 & 158.937 & 57.188 & $1.08 \pm 0.15$ & 1.27 (spec) \\
83 & 2606 & SRGe J103512.3+575548 & 2 & $0.12^{+0.44}_{-0.43}$ & $1.40^{+1.61}_{-1.11}$ & $2.08^{+2.83}_{-1.67}$ & 158.801 & 57.930 & 158.801 & 57.930 & $0.99 \pm 0.15$ & 0.72 (spec) \\
84 & 2633 & SRGe J103728.8+581948 & 2 & $0.13^{+0.43}_{-0.41}$ & $2.10^{+2.35}_{-1.33}$ & $1.80^{+2.04}_{-1.13}$ & 159.370 & 58.330 & 159.370 & 58.330 & $0.98 \pm 0.15$ & 0.54 (phot) \\
85 & 2875 & SRGe J104532.9+562403 & 2 & $0.60^{+0.61}_{-0.59}$ & $1.70^{+1.89}_{-1.27}$ & $1.09^{+2.83}_{-1.07}$ & 161.387 & 56.401 & 161.387 & 56.401 & $0.91 \pm 0.15$ & 0.83 (phot) \\
86 & 3150 & SRGe J102210.6+564525 & 2 & $0.58^{+0.70}_{-0.67}$ & $ < 5.30 $ & $8.80^{+15.33}_{-6.16}$ & 155.544 & 56.757 & 155.542 & 56.757 & $0.83 \pm 0.16$ & 1.13 (spec) \\
$\vdots$ & $\vdots$ & $\vdots$ & $\vdots$ & $\vdots$ & $\vdots$ & $\vdots$ & $\vdots$ & $\vdots$ & $\vdots$ & $\vdots$ & $\vdots$ & $\vdots$ \\
286 & 529 & SRGe J104648.6+541225 & 3 & $-0.08^{+1.21}_{-1.19}$ & $-$ & $-$ & 161.702 & 54.207 & 161.701 & 54.208 & $3.37 \pm 0.84$ & $-$ \\
287 & 831 & SRGe J103415.3+595245 & 3 & $0.45^{+0.56}_{-0.53}$ & $-$ & $-$ & 158.564 & 59.879 & 158.563 & 59.880 & $2.46 \pm 0.39$ & $-$ \\
288 & 1547 & SRGe J105141.8+584025 & 3 & $0.72^{+0.40}_{-0.38}$ & $-$ & $-$ & 162.924 & 58.673 & 162.925 & 58.673 & $1.51 \pm 0.19$ & $-$ \\
289 & 1578 & SRGe J105542.1+595218 & 3 & $0.30^{+0.56}_{-0.52}$ & $-$ & $-$ & 163.925 & 59.872 & 163.925 & 59.871 & $1.49 \pm 0.24$ & $-$ \\
290 & 1600 & SRGe J104633.4+573928 & 3 & $0.87^{+0.41}_{-0.40}$ & $-$ & $-$ & 161.639 & 57.658 & 161.639 & 57.658 & $1.47 \pm 0.19$ & $-$ \\
291 & 1820 & SRGe J104154.3+555258 & 3 & $0.73^{+0.44}_{-0.42}$ & $-$ & $-$ & 160.476 & 55.883 & 160.477 & 55.883 & $1.33 \pm 0.17$ & $-$ \\
\bottomrule
\caption{\srge CT AGN candidates in \lh, faint source catalog. Columns are the same as in Table \ref{tab:CT_AGN_candidates_bright}. The full table is available in supplementary materials.}
% \end{tabularx}
\end{longtable}
    \label{tab:CT_AGN_candidates_faint}
\end{landscape}
\twocolumn

% Don't change these lines
\bsp	% typesetting comment
\label{lastpage}
\end{document}